\documentclass{webofc}
\usepackage[varg]{txfonts}   
\begin{document}
\title{NIKA: a mm camera for Sunyaev-Zel'dovich science in clusters of galaxies}
%
%

\author{\firstname{J.F.} \lastname{Mac\'ias-P\'erez} \inst{\ref{LPSC}}\fnsep\thanks{\email{macias@lpsc.in2p3.fr}}
\and \firstname{R.} \lastname{Adam} \inst{\ref{LLR},\ref{CEFCA}}
\and  \firstname{P.} \lastname{Ade} \inst{\ref{Cardiff}}
\and  \firstname{P.} \lastname{Andr\'e} \inst{\ref{CEA1}}
\and  \firstname{A.} \lastname{Andrianasolo} \inst{\ref{IPAG}}
\and  \firstname{H.} \lastname{Aussel} \inst{\ref{CEA1}}
\and  \firstname{M.} \lastname{Arnaud} \inst{\ref{CEA1}}
\and  \firstname{I.} \lastname{Bartalucci} \inst{\ref{CEA1}}
\and  \firstname{A.} \lastname{Beelen} \inst{\ref{IAS}}
\and  \firstname{A.} \lastname{Beno\^it} \inst{\ref{Neel}}
\and  \firstname{A.} \lastname{Bideaud} \inst{\ref{Neel}}
\and  \firstname{O.} \lastname{Bourrion} \inst{\ref{LPSC}}
\and  \firstname{M.} \lastname{Calvo} \inst{\ref{Neel}}
\and  \firstname{A.} \lastname{Catalano} \inst{\ref{LPSC}}
\and  \firstname{B.} \lastname{Comis} \inst{\ref{LPSC}}
\and  \firstname{M.} \lastname{De~Petris} \inst{\ref{Roma}}
\and  \firstname{F.-X.} \lastname{D\'esert} \inst{\ref{IPAG}}
\and  \firstname{S.} \lastname{Doyle} \inst{\ref{Cardiff}}
\and  \firstname{E.~F.~C.} \lastname{Driessen} \inst{\ref{IRAMF}}
\and  \firstname{A.} \lastname{Gomez} \inst{\ref{CAB}}
\and  \firstname{J.} \lastname{Goupy} \inst{\ref{Neel}}
\and  \firstname{F.} \lastname{K\'eruzor\'e} \inst{\ref{LPSC}}
\and  \firstname{C.} \lastname{Kramer} \inst{\ref{IRAME}}
\and  \firstname{B.} \lastname{Ladjelate} \inst{\ref{IRAME}}
\and  \firstname{G.} \lastname{Lagache} \inst{\ref{LAM}}
\and  \firstname{S.} \lastname{Leclercq} \inst{\ref{IRAMF}}
\and  \firstname{J.-F.} \lastname{Lestrade} \inst{\ref{LERMA}}
\and  \firstname{P.} \lastname{Mauskopf} \inst{\ref{Cardiff},\ref{Arizona}}
\and \firstname{F.} \lastname{Mayet} \inst{\ref{LPSC}}
\and  \firstname{A.} \lastname{Monfardini} \inst{\ref{Neel}}
\and  \firstname{L.} \lastname{Perotto} \inst{\ref{LPSC}}
\and  \firstname{G.} \lastname{Pisano} \inst{\ref{Cardiff}}
\and  \firstname{E.} \lastname{Pointecouteau} \inst{\ref{IRAP}}
\and  \firstname{N.} \lastname{Ponthieu} \inst{\ref{IPAG}}
\and  \firstname{G.W.} \lastname{Pratt} \inst{\ref{CEA1}}
\and  \firstname{V.} \lastname{Rev\'eret} \inst{\ref{CEA1}}
\and  \firstname{A.} \lastname{Ritacco} \inst{\ref{IRAME}}
\and  \firstname{C.} \lastname{Romero} \inst{\ref{IRAMF}}
\and  \firstname{H.} \lastname{Roussel} \inst{\ref{IAP}}
\and  \firstname{F.} \lastname{Ruppin} \inst{\ref{MIT}}
\and  \firstname{K.} \lastname{Schuster} \inst{\ref{IRAMF}}
\and  \firstname{S.} \lastname{Shu} \inst{\ref{IRAMF}}
\and  \firstname{A.} \lastname{Sievers} \inst{\ref{IRAME}}
\and  \firstname{C.} \lastname{Tucker} \inst{\ref{Cardiff}}
\and  \firstname{R.} \lastname{Zylka} \inst{\ref{IRAMF}}
}

\institute{\label{LPSC} Univ. Grenoble Alpes, CNRS, Grenoble INP, LPSC-IN2P3, 53, avenue des Martyrs, 38000 Grenoble, France
\and \label{LLR} LLR (Laboratoire Leprince-Ringuet), CNRS, \'Ecole Polytechnique, Institut Polytechnique de Paris, Palaiseau, France  
\and \label{CEFCA} Centro de Estudios de F\'isica del Cosmos de Arag\'on (CEFCA), Plaza San Juan, 1, planta 2, E-44001, Teruel, Spain 
\and \label{Cardiff} Astronomy Instrumentation Group, University of Cardiff, UK          
\and \label{CEA1} AIM, CEA, CNRS, Universit\'e Paris-Saclay, Universit\'e Paris Diderot, Sorbonne Paris Cit\'e, 91191 Gif-sur-Yvette, France     
\and \label{IPAG} Univ. Grenoble Alpes, CNRS, IPAG, 38000 Grenoble, France     
\and \label{IAS} Institut d'Astrophysique Spatiale (IAS), CNRS and Universit\'e Paris Sud, Orsay, France    
\and \label{Neel} Institut N\'eel, CNRS and Universit\'e Grenoble Alpes, France
\and \label{Roma} Dipartimento di Fisica, Sapienza Universit\`a di Roma, Piazzale Aldo Moro 5, I-00185 Roma, Italy       
\and \label{IRAMF} Institut de RadioAstronomie Millim\'etrique (IRAM), Grenoble, France 
\and \label{CAB} Centro de Astrobiolog\'ia (CSIC-INTA), Torrej\'on de Ardoz, 28850 Madrid, Spain
\and \label{IRAME} Instituto de Radioastronom\'ia Milim\'etrica (IRAM), Granada, Spain 
\and \label{LAM} Aix Marseille Univ, CNRS, CNES, LAM (Laboratoire d'Astrophysique de Marseille), Marseille, France
\and \label{LERMA} LERMA, Observatoire de Paris, PSL Research University, CNRS, Sorbonne Universit\'es, UPMC Univ. Paris 06, 75014 Paris,
France
\and \label{Arizona} School of Earth and Space Exploration and Department of Physics, Arizona State University, Tempe, AZ 85287         
\and \label{IAP} Institut d'Astrophysique de Paris, CNRS (UMR7095), 98 bis boulevard Arago, 75014 Paris, France
\and \label{IRAP} IRAP, Universit\'e de Toulouse, CNRS, CNES, UPS, (Toulouse), France
\and \label{MIT} Kavli Institute for Astrophysics and Space Research, Massachusetts Institute of Technology, Cambridge, MA 02139, USA 
          }

\abstract{%
 Clusters of galaxies, the largest bound objects in the Universe, constitute a cosmological probe of choice, which is sensitive to both dark matter and dark energy. Within this framework, the Sunyaev-Zel'dovich (SZ) effect has opened a new window for the detection of clusters of galaxies and for the characterization of their physical properties such as mass, pressure and temperature. NIKA, a KID-based dual band camera installed at the IRAM 30-m telescope, was particularly well adapted in terms of frequency, angular resolution, field-of-view and sensitivity, for the mapping of the thermal and kinetic SZ effect in high-redshift clusters. In this paper, we present the NIKA cluster sample and a review of the main results obtained via the measurement of the SZ effect on those clusters: reconstruction of the cluster radial pressure profile, mass, temperature and velocity.
}
\maketitle
\section{Introduction}
\label{sec:intro}

\begin{figure}[h]
\begin{center}
\sidecaption
\includegraphics[scale=0.3]{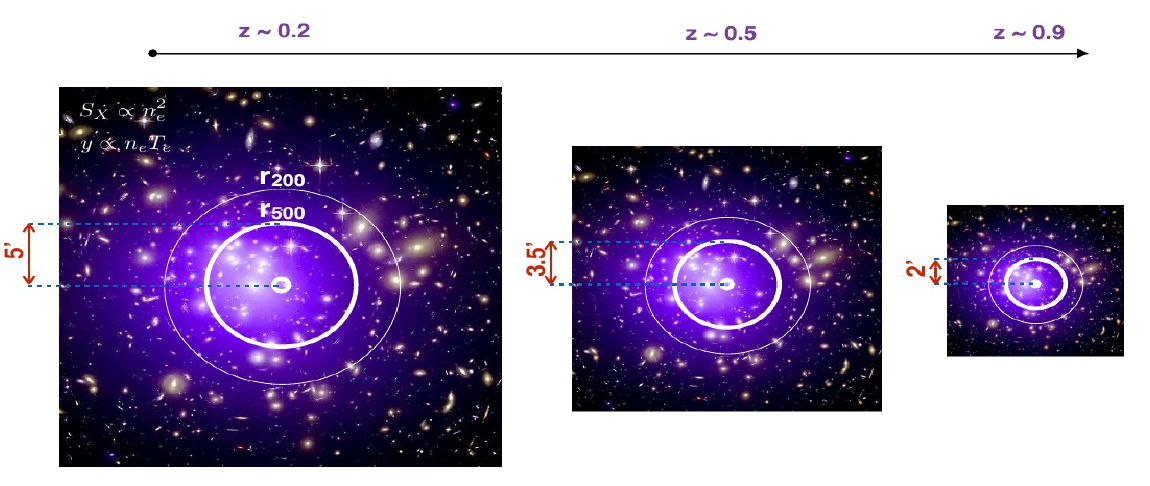}
\label{fig-mclusters}       
\caption{Evolution of cluster angular size with redshift for fixed mass.}
\end{center}
\end{figure}
Clusters of galaxies are the largest gravitationally bound objects in the Universe and trace the matter distribution across cosmological times \cite{voit2005}. Clusters are mainly made of dark matter (about 85\% of their total mass), but also of hot ionized gas (about 12\%) and of the stars and interstellar medium within galaxies (a few percent). The latter two represent the baryonic component of clusters and can be used to detect and study them. At visible and IR wavelengths we can observe cluster galaxies, while the baryonic hot ionized gas that form the Intra Cluster Medium (ICM) can be detected via its X-ray emission \cite{boehringer} and the thermal Sunyaev-Zeldovich (tSZ) effect \cite{sz}. The latter corresponds to the Compton inverse interaction of the ICM electrons with the CMB photons travelling through the cluster, and leads to a well defined spectral distortion of the CMB emission.\\

\noindent Clusters are a powerful probe for cosmology because they form across the expansion of the Universe (see for example \cite{Allen}). Their abundance as a function of mass and redshift is very sensitive to cosmological parameters \cite{PlanckNC} such as $\sigma_{8}$ (the rms of the matter perturbations at 8 Mpc scales), $\Omega_{m}$ (the dark matter density), $\Omega_{\Lambda}$ (the dark energy density). Constraints on these parameters derived from galaxy cluster samples are generally limited by the accuracy of mass estimates of galaxy clusters \cite{hasselfield2013,dehaan2016}, which mainly come from the baryonic observables. Scaling relations relating the mass of the cluster to the baryonic observables are generally used (see \cite{pratt2019} for a review). These scaling relations assume that clusters are relaxed and that gravity is the only physics at play. Furthermore, they are generally calibrated using low redshift clusters. However, baryonic physics -- for example shocks during merging events, turbulence in the gas, and cooling-flows near active galactic nuclei -- may introduce deviations with respect to the self-similar scenario and lead to significant bias in the cluster mass estimates. Such deviations are expected to be more likely at high redshift as merging processes are expected to be more common following the hierarchical scenario of structure formation. \\

Within the self-similar scenario cluster properties are only linked to their mass and redshift. In particular, as illustrated in Figure~\ref{fig-mclusters}, for an equivalent mass clusters in the redshift range $0.5 < z < 1.0$ are observed smaller in angular size than the low redshift ones, $z<0.3$. Therefore, a better understanding of high redshift cluster properties can only be achieved via high resolution tSZ observations \cite{2019SSRv..215...17M}. The NIKA camera (\cite{2011ApJS..194...24M}), installed at the 30 m IRAM telescope in Pico Veleta, was a pioneer in this respect as will be shown in this paper.

\section{The NIKA camera}
\label{sec:nika}

\begin{figure}[h]
\begin{center}
\sidecaption
\includegraphics[scale=0.3]{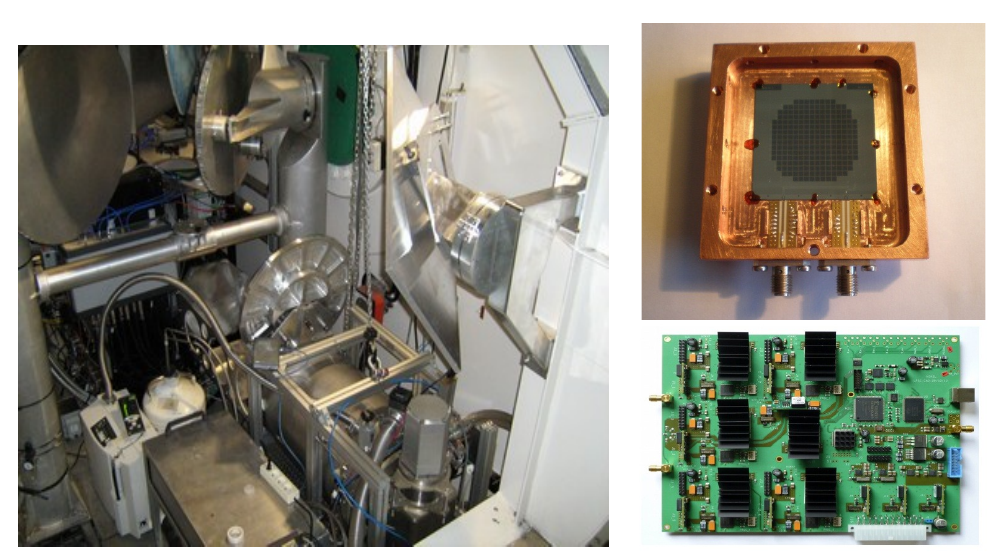}
\label{fig-nikainstru}       
\caption{The NIKA instrument: cryostat installed at the 30 m telescope cabin, array of KIDs and readout electronic.}
\end{center}
\end{figure}

\begin{table}
\centering
\caption{NIKA instrumental performance in intensity \cite{Catalano:2014nml}. \label{tab-performance}       
}
\begin{tabular}{lll}
\hline
       &  150 GHz & 260 GHz  \\
\hline
Number of KIDs & 132 & 224\\
FOV [arcmin] & 1.8 &  2.0\\
Sensitivity [mJy/s$^{1/2}$] & 14  & 40 \\
Resolution [arcsec] & 18 & 12 \\
\hline
\end{tabular}
\end{table}

NIKA \cite{2011ApJS..194...24M, 2012JLTP..167..834M} was a dual band millimeter intensity and polarization camera operated at 150 and 260 GHz and installed permanently at the IRAM 30-m telescope from 2013 to 2015. The NIKA camera (see Figure~\ref{fig-nikainstru}) was made of two arrays of Kinetic Inductance Detectors (KIDs) cooled down to 100 mK via a $^3$He-$^4$He dilution cryostat and instrumented via a dedicated readout electronics \cite{2012JInst...7.7014B, 2012SPIE.8452E..0OB}. NIKA has been the first KID-based camera to produce scientific quality results \cite{2014A&A...569A..66A} and has demonstrated state of the art performance during operations \cite{2013A&A...551L..12C,Catalano:2014nml}. Table~\ref{tab-performance} summarizes the main characteristic of the NIKA camera and its performance during operations.

\section{The NIKA cluster sample}
\label{sec:clustersample}

\begin{figure}[h]
\begin{center}
\includegraphics[scale=0.45]{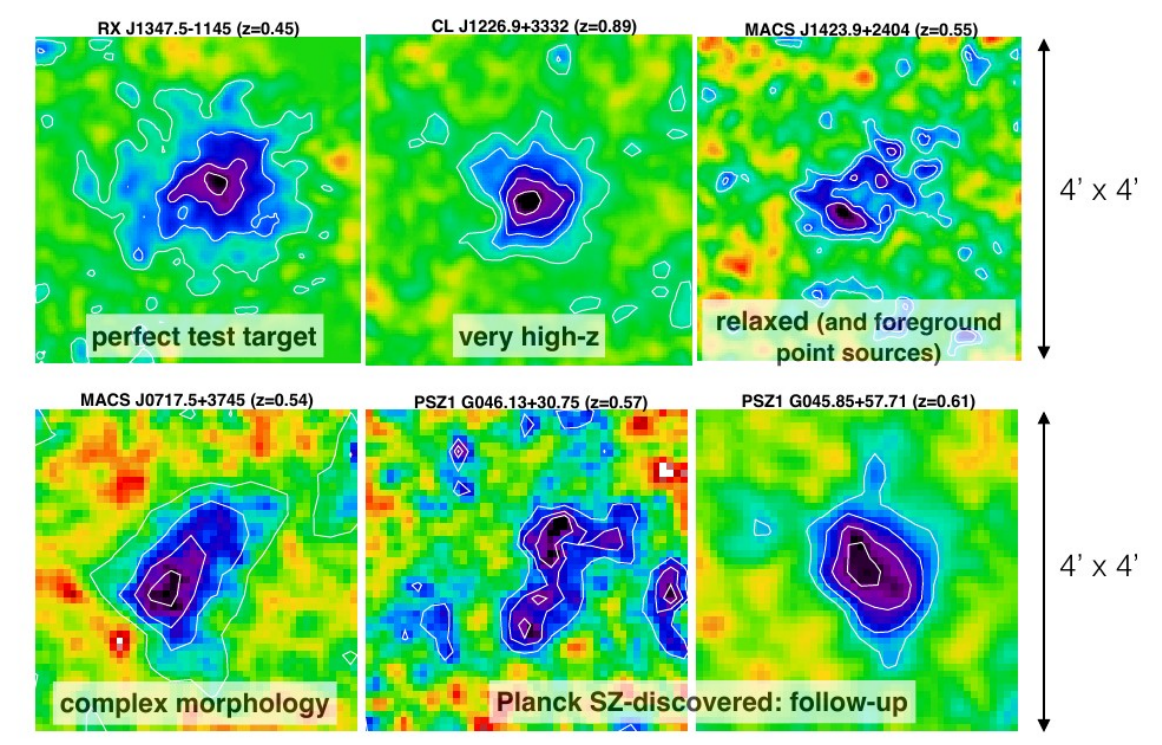}
\label{fig-clustersample}       
\caption{The NIKA cluster sample. NIKA maps of the tSZ effect at 150~GHz for the clusters: \textsc{RXJ1347.5-1145}, \textsc{CLJ1216.9+3332},
\textsc{MACS J1423.9+2404}, \textsc{MACS J0717.5+3745},\textsc{PSZ1 G045.85+57.71} and \textsc{PSZ1 G046.13+30.75}.}
\end{center}
\end{figure}

The NIKA camera was particularly well adapted for observations of the SZ effect in clusters of galaxies at high redshift because of: 1) the dual band capabilities with frequencies sampling the zero (260~GHz) and negative part (150~GHz) of the thermal SZ spectrum \cite{sz}, 2) the high resolution offered by the 30 m telescope and the large FOV, which permitted a detailed mapping of clusters in the redshift range from 0.5 to 1, 3) excellent performance in sensitivity allowing fast mapping speed, and 4) an accurate calibration and photometry. \\

Because of this, during NIKA operations it has been possible to map a sample of 6 clusters of galaxies\footnote{\url{http://lpsc.in2p3.fr/NIKA2LPSZ/nika2sz.release.php}}
 as shown in Figure~\ref{fig-clustersample}. The cluster sample was chosen in order to best explore the capabilities of large KID-based cameras for cluster science using the SZ effect. The first cluster observed was \textsc{RXJ1347.5-1145} \cite{2014A&A...569A..66A}, which is a very massive and medium redshift, z=0.45, cluster and constitutes a perfect first target. These observations were the first ever scientific quality observations with a KID camera. To further test the capabilities of NIKA, there were observations of \textsc{CLJ1216.9+3332}, which is a massive and high redshift cluster, z=0.89 \cite{2015A&A...576A..12A,romero}. One important issue with the observations of high redshift cluster via the tSZ effect is the contamination by dusty and radio point-like sources as was shown by the NIKA maps of the cluster \textsc{MACS J1423.9+2404} presented in \cite{2016A&A...586A.122A}. The high resolution and large FOV capabilities of NIKA allowed also the detailed study of \textsc{MACS J0717.5+3745} \cite{2017A&A...598A.115A,2017A&A...606A..64A,2018A&A...614A.118A}, which is a complex morphology cluster presenting various components as well as extreme physical conditions (violent merging events, large velocities, etc). Finally, it was possible to check that the follow-up of high redshift clusters detected (e.g. \textsc{PSZ1 G045.85+57.71} and \textsc{PSZ1 G046.13+30.75} ) by low resolution CMB experiments like Planck is possible with NIKA like cameras. This work demonstrated that cluster pressure profile and  mass estimates can be significantly improved as in the case of \textsc{PSZ1 G045.85+57.71} \cite{2017A&A...597A.110R}.

\section{NIKA results on SZ science}
\label{sec:nikasz}

The NIKA camera has permitted a wide sample of SZ studies on clusters of galaxies. Here we have selected some representative examples.

\subsection{Cluster pressure profile estimation}
\label{sec-pressure}

\begin{figure}[h]
\begin{center}
\includegraphics[scale=0.4]{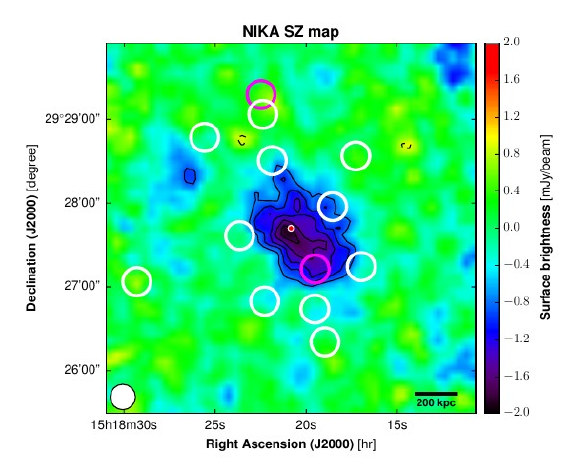}
\includegraphics[scale=0.33]{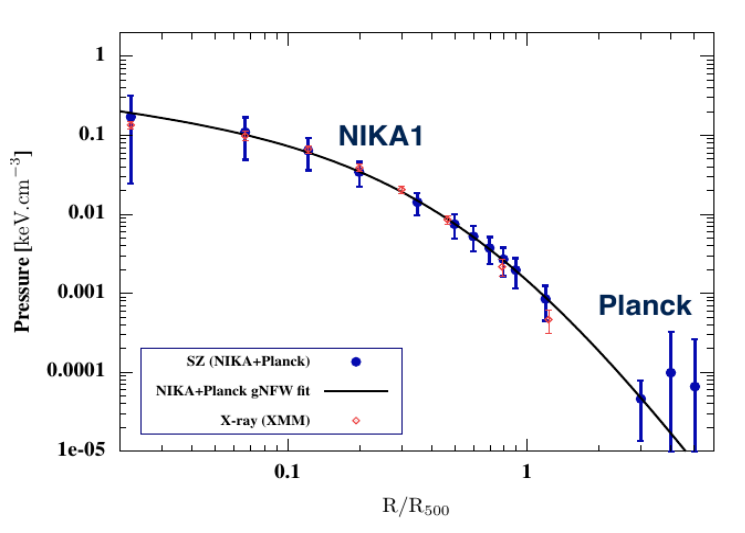}
\caption{Pressure profile reconstruction for the cluster \textsc{PSZ1 G045.85+57.71} \cite{2017A&A...597A.110R}. Left: thermal SZ map of the cluster at 150~GHz. Right: Non-parametric reconstruction of the cluster pressure profile as a function of radius as obtained from the NIKA and Planck data.\protect\footnotemark \label{fig-clusterprofile} }
\end{center}
\end{figure}
\footnotetext{$R_{500}$ is defined as the radius at which the mean cluster density is 500 times the cosmological critical density.} 

Cosmological analyses with cluster of galaxies (e.g. \cite{PlanckNC}) require accurate measurements of the cluster mass. This can be achieved using the SZ effect only via scaling relations between the cluster mass and the integrated Compton parameter \cite{pratt2019} or from a combination of the SZ and X-ray data by computing the cluster hydrostatic mass. In both cases the detailed reconstruction of the cluster pressure profile is a key element. NIKA has shown that this can be achieved with high resolution KIDs-based cameras both for parametric \cite{2015A&A...576A..12A,2016A&A...586A.122A} and non-parametric models \cite{romero,2017A&A...597A.110R}. The latter case is illustrated in Figure~\ref{fig-clusterprofile}. As observed in the right panel of the figure, the combination of the high-resolution and high-sensitivity NIKA data with the Planck data permitted the non-parametric reconstruction of the pressure profile for the cluster \textsc{PSZ1 G045.85+57.71} from the inner part of the cluster to the outskirts.

\subsection{Cluster velocity}
\label{sec-pressure}
 
 \begin{figure}[h]
\begin{center}
\includegraphics[scale=0.32]{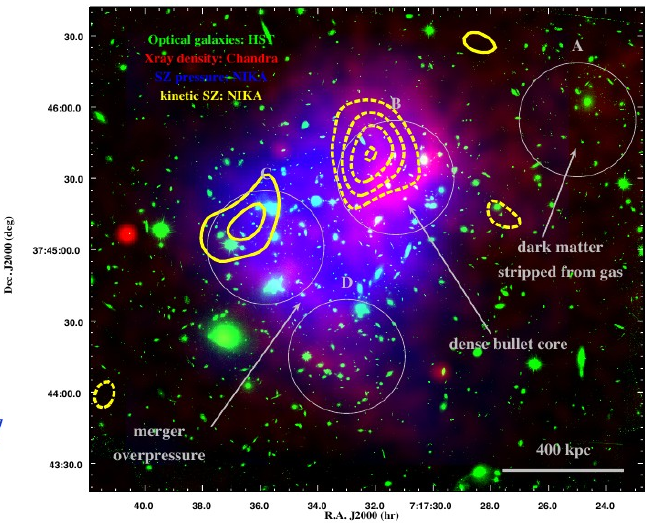}
\includegraphics[scale=0.34,trim=0 0 0.55cm 0, clip]{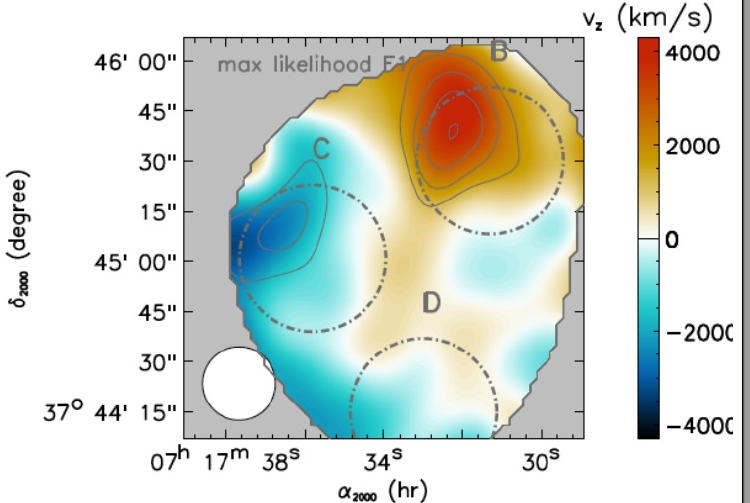}
\caption{Kinetic SZ effect observed by NIKA in the cluster \textsc{MACS J0717.5+3745} \label{fig-clustervelocity} }
\end{center}
\end{figure}

The kinetic SZ effect \cite{ksz}, arising from the CMB Doppler shift produced by the bulk motion of the ICM electrons,
can be used to measure the velocity of cluster of galaxies along the line-of-sight. By contrast to the thermal SZ effect, the kinetic SZ effect does not produce a spectral distortion of the CMB photons and it shows the same spectrum that the CMB. Thus, in the case of the NIKA observations we expect to observe the same signal in CMB temperature units at 150 and 260~GHz. Furthermore, the kinetic SZ effect is expected to be small with respect to the thermal one for typical clusters velocities, which are typically in the range of few hundreds to few thousands of km/s \cite{2013ApJ...778...52S,2016ApJ...820..101S,2017A&A...598A.115A,2019ApJ...880...45S}. In this respect, the cluster \textsc{MACS J0717.5+3745} is a target of choice as we expect the different components in the cluster to present large relative velocity differences \cite{2013ApJ...778...52S,2017A&A...598A.115A}. This is illustrated in Figure~\ref{fig-clustervelocity}. \\

In the right panel of the figure we show a composite map of \textsc{MACS J0717.5+3745} obtained from observations at various wavelengths: optical image (green), X-ray (red), and thermal (blue) and kinetic (yellow) SZ as measured by NIKA \cite{2017A&A...598A.115A}. These latter maps are obtained from the combination of the NIKA 150 and 260~GHz maps after cleaning for astrophysical contaminants and accounting for temperature induced relativistic corrections \cite{2017A&A...598A.115A}. The different subclusters present in \textsc{MACS J0717.5+3745} are shown as dotted-dashed circles. From the kinetic SZ map, it is possible to extract a velocity map that is shown on the right panel of Figure~\ref{fig-clustervelocity}. We observe in this map, the first model independent one, that the substructures C and D have very large velocities along the line-of-sight, with opposite sign.

\subsection{Cluster temperature estimation}
\label{sec-pressure}
 
\begin{figure}[h]
\begin{center}
\includegraphics[scale=0.38]{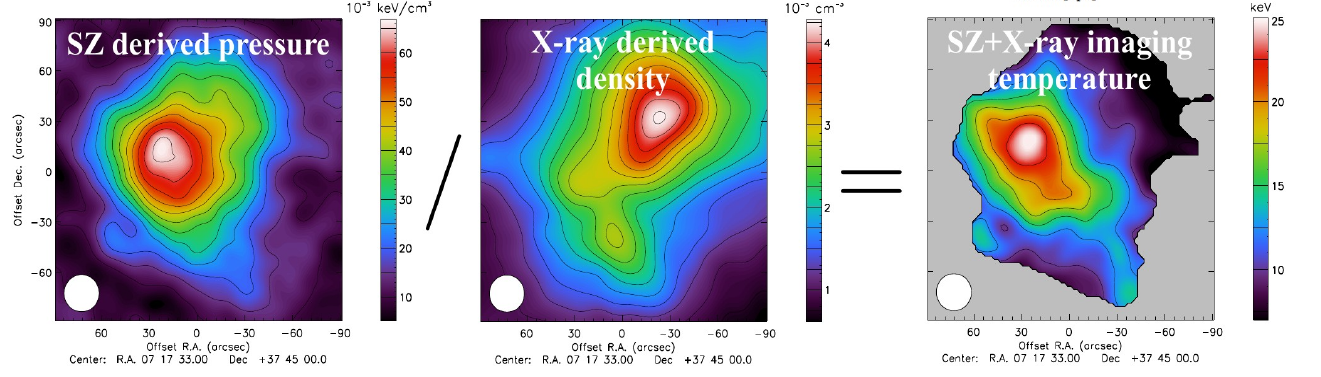}
\caption{Reconstruction of the projected temperature of the cluster \textsc{MACS J0717.5+3745} from a combined SZ and X-ray analysis. The maps of the projected electron pressure, density and temperature are shown from left to right. \label{fig-clustemperature} }
\end{center}
\end{figure}

The cluster temperature can be generally estimated from X-ray spectroscopic observations. However, these measurements are affected by 
two major systematics: 1) the X-ray brightness temperature is proportional to the square of the electron density, such that the spectroscopic
temperatures are driven by the colder and denser regions along the line-of-sight  \cite{mazzotta2004}, and 2) the recovered temperature estimates are very sensitive to the calibration scale of current X-ray satelite observatories leading to absolute uncertainties of about 15 \% (e.g. \cite{mahdavi2013}). Furthermore, spectroscopic observations are very time costly. Alternatively, the SZ effect can also be used to measure cluster temperature given a measurement of the electron density using for example X-ray photometric observations. This type of analysis was performed for the first time by \cite{2017A&A...606A..64A} for the cluster \textsc{MACS J0717.5+3745} using a 2D reconstruction of the electron pressure from the NIKA SZ data accounting for relativistic corrections as in \cite{2017A&A...598A.115A}, and of the 2D electron density derived from the XMM-Newton X-ray data. This is illustrated in Figure~\ref{fig-clustemperature}, which presents the maps of the electron pressure, density and reconstructed temperature. We observe a high signal-to-noise reconstruction of the cluster, which shows huge increase at the position of the shock between two of the substructures in the cluster. We find that the combined SZ and X-ray based temperature map is compatible with those obtained from X-ray spectroscopy using XMM-Newton and Chandra observations to 10 \%  in amplitude after correcting for the kinetic SZ effect. This is a promising result and opens a new window for temperature determination in clusters of galaxies. 

\section{Conclusions}
\label{sec:conclu}
The NIKA camera was demonstrated to be an excellent instrument for the study of clusters of galaxies
via the Sunyaev-Zel'dovich effect. Among others, NIKA has permitted the detailed characterization of a sample of six galaxy clusters, chosen to demonstrate the capabilities of KID-based instruments.
NIKA has provided the first ever SZ scientific observations with KIDs and permitted to conduct a wide scientific program. This program included for example the reconstruction of the cluster pressure profile, temperature and mass via the thermal SZ effect in combination with X-ray observations, as well as the first direct mapping of cluster velocity via the kinetic SZ effect.

\section*{Acknowledgements}
\begin{scriptsize}
We would like to thank the IRAM staff for their support during the campaigns. The NIKA dilution cryostat has been designed and built at the Institut N\'eel. In particular, we acknowledge the crucial contribution of the Cryogenics Group, and in particular Gregory Garde, Henri Rodenas, Jean Paul Leggeri, Philippe Camus. This work has been partially funded by the Foundation Nanoscience Grenoble and the LabEx FOCUS ANR-11-LABX-0013. This work is supported by the French National Research Agency under the contracts "MKIDS", "NIKA" and ANR-15-CE31-0017 and in the framework of the "Investissements d’avenir” program (ANR-15-IDEX-02). This work has benefited from the support of the European Research Council Advanced Grant ORISTARS under the European Union's Seventh Framework Programme (Grant Agreement no. 291294). We acknowledge fundings from the ENIGMASS French LabEx (R. A. and F. R.), the CNES post-doctoral fellowship program (R. A.), the CNES doctoral fellowship program (A. R.) and the FOCUS French LabEx doctoral fellowship program (A. R.). R.A. acknowledges support from Spanish Ministerio de Econom\'ia and Competitividad (MINECO) through grant number AYA2015-66211-C2-2.

\end{scriptsize}
%
%
%
\begin{scriptsize}

\end{scriptsize}

\end{document}